# A Cryogenic Readout IC with 100 KSPS in-Pixel ADC for Skipper CCD-in-CMOS Sensors

Adam Quinn, Manuel B. Valentin, Thomas Zimmerman, Davide Braga, Seda Memik, Farah Fahim

*Abstract*—The Skipper CCD-in-CMOS Parallel Read-Out Circuit (SPROCKET) is a mixed-signal front-end design for the readout of Skipper CCD-in-CMOS image sensors. SPROCKET is fabricated in a 65 nm CMOS process and each pixel occupies a 50 µm × 50 µm footprint. SPROCKET is intended to be heterogeneously integrated with a Skipper-in-CMOS sensor array, such that one readout pixel is connected to a multiplexed array of nine Skipper-in-CMOS pixels to enable massively parallel readout. The front-end includes a variable gain preamplifier, a correlated double sampling circuit, and a 10-bit serial successive approximation register (SAR) ADC. The circuit achieves a sample rate of 100 ksps with 0.48 $e^-_{rms}$ equivalent noise at the input to the ADC. SPROCKET achieves a maximum dynamic range of 9,000 $e^-$ at the lowest gain setting (or 900 $e^-$ at the lowest noise setting). The circuit operates at 100 Kelvin with a power consumption of 40 *µW* per pixel. A SPROCKET test chip was submitted in September 2022, and test results will be presented at the conference.

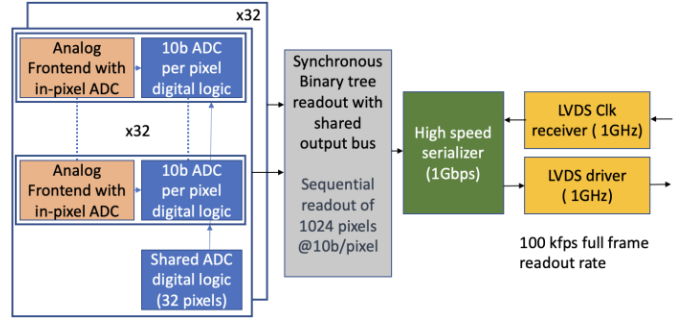

Fig. 1. Functional block diagram of one SPROCKET mini-array consisting of 1024 pixels, including readout by a synchronous binary tree [9]

## I. INTRODUCTION

Future high energy physics (HEP) and dark matter detection experiments [1], as well as quantum imaging applications, will continue to require extremely sensitive and low-noise particle detectors with larger area and thus higher data throughput. Charge-Coupled Device (CCD) cameras offer excellent performance in a range of scientific imaging applications [2] [3]. In particular, Skipper-CCDs, and the Skipper-in-CMOS sensor currently being developed by Fermilab and SLAC, are a highly attractive class of detectors because they allow sub-single-electron noise to be achieved by reading out the same packet of charge many times [4] [5] [6]. However, traditional monolithic architectures in which charge is read out only from the edge of the CCD array cannot scale to megapixel array sizes without severely compromising readout speed. A promising alternative is offered by "hybrid" readout architectures which integrate a readout integrated circuit (ROIC) directly beneath the image sensor, thus allowing each Skipper CCD-in-CMOS pixel (or a multiplexed group of several pixels) to be bump bonded to a dedicated readout circuit (see Fig. 2).

Integrating readout electronics per-pixel imposes severe constraints on the design of the analog front-end [7]. The front-end must be compact and consume low static and dynamic power, particularly if operated at cryogenic temperatures. These constraints must be met without compromising the noise performance of the detector. To trade between area and readout speed, SPROCKET is designed to interface with a multiplexed block of sixteen pixels. This allows a readout circuit area of 60 µm × 60 µm for a pixel footprint of 15 µm × 15 µm.

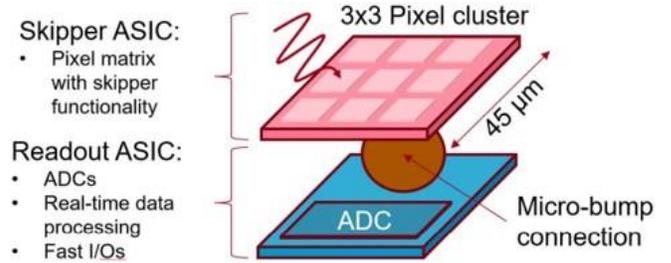

Fig. 2. Illustration of SPROCKET's hybrid integration approach

The first SPROCKET test chip, which is based on the design described in [8] with modifications to increase sampling rate by a factor of three, includes two 32x32 mini-arrays of pixels with per-array high-speed readout [9], as well as a test structure for de-embedding noise performance. This test chip features a pixel area of 50 µm × 50 µm; this will be expanded to 60 µm × 60 µm for the final tapeout to enable bump bonding to the Skipper-CCD-in-CMOS ASIC. The mini-array architecture is shown in Fig. 1. The following sections describe the pixel design architecture and simulation results, a brief discussion of array-level integration, and directions for future work.

## II. DESIGN ARCHITECTURE

Fig. 3 shows a schematic of SPROCKET. When the ROIC is bump bonded to a Skipper CCD-in-CMOS sensor, the node labeled "Bump Bond" will be connected to the source of the integrated source follower in the sensor's output stage. A bias current of approximately 5 µA will flow from the integrated source follower through the bump bond and into the NMOS termination shown in SPROCKET, such that the input to SPROCKET is a voltage-mode signal. The three stages of the SPROCKET front-end (preamplifier, correlated double-sampling circuit, and ADC) are presented in the following sections in the order of signal processing.

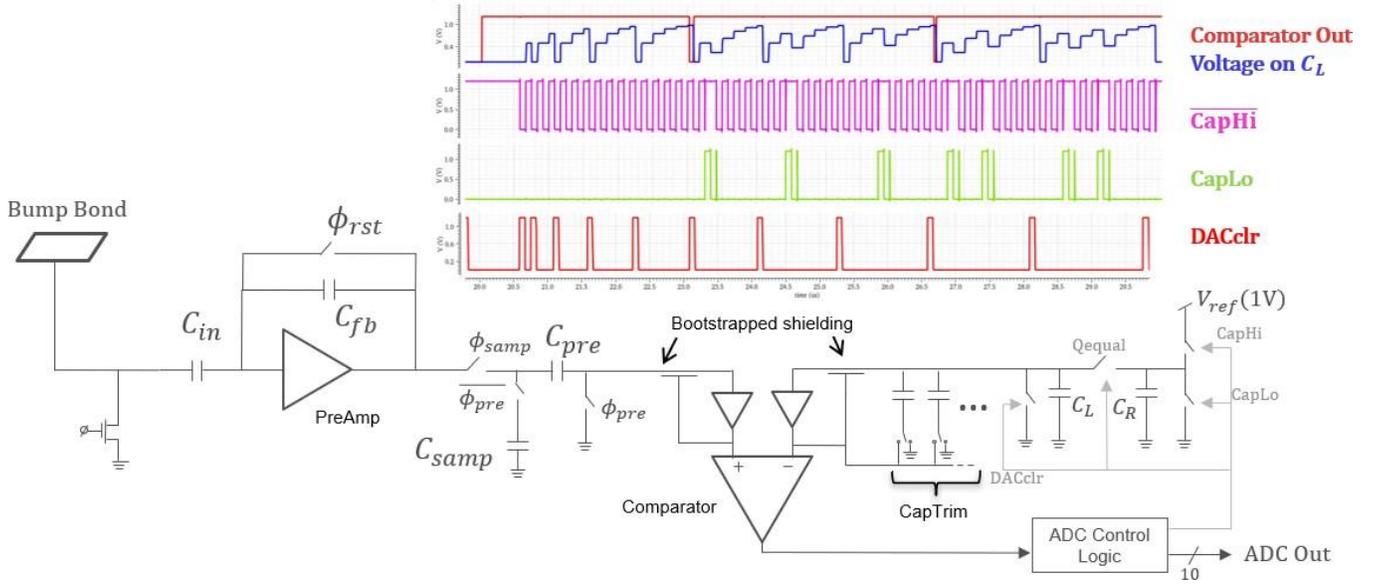

Fig. 3. Simplified schematic of SPROCKET, modified from [8], with example waveforms from one 10-bit ADC acquisition.

### A. Preamplifier

A preamplifier is placed immediately at the input of SPROCKET in order to maximize noise figure. The core of the amplifier is a gain-boosted cascode, and its gain is determined by the capacitive feedback network composed of $C_{in}$ and $C_{fb}$. A reset switch provides a bias to the high-impedance node at the input to the amplifier. $C_{in}$ and $C_{fb}$ are implemented as identical 3 fF metal-oxide-metal (MOM) capacitors, giving approximately unity gain.

The preamplifier has two special modes of operation. In *high-gain mode*, an additional 27 fF capacitor $C_{range2}$ (not shown) is switched in parallel with $C_{in}$. This increases the gain of the preamplifier to 10x. In *inject mode*, the bump bond node is directly connected to the global *Inject* net, which allows the injection of a test signal.

### B. Correlated Double-Sampling Circuit

The preamplifier is immediately followed by a correlated double-sampling (CDS) circuit composed of two sampling capacitors $C_{pre}$ and $C_{samp}$ and their associated switches.

In the Pre-Sample phase, the CCD pedestal level is stored across $C_{pre}$, and in the Sample phase the signal (including pedestal) is stored across $C_{samp}$. The resulting voltage at the comparator input is:

$$V(C_{samp}) - V(C_{pre}) = (V_{sig} + V_{ped}) - V_{ped} = V_{sig}$$

The complement switch $\overline{\phi_{pre}}$ disconnects $C_{samp}$ from the output of the preamplifier during the Pre-Sample phase. This switch ensures that in the Pre-Sample phase, the preamplifier sees only the capacitive load of $C_{pre}$, rather than $C_{pre}$ + $C_{samp}$. Both $C_{samp}$ and $C_{pre}$ are relatively large (> 100 fF), so this techniques avoids over-design in the preamplifier and ensures better pedestal cancellation by presenting similar impedances to the preamplifier in both phases [10].

The sample and pre-sample switches are implemented as transmission gates to mitigate charge injection. Residual charge injection contributes an estimated < 0.5 mV non-linearity over the dynamic range of the ADC.

### C. Zero Input-Capacitance Comparator

The sampled voltage serves as the input to a zero input-capacitance comparator, first presented in [11], which forms the core of the SPROCKET ADC. The novel contribution of this comparator is the use of input buffers to generate level-shifted versions of the comparator's positive and negative inputs, *OutBufp* and *OutBufm*. These buffered signals can be used to shield the input nodes as shown in Fig. 3 because any capacitance between the input node and its corresponding buffered signal is bootstrapped. This technique has two major benefits: First, by tying the well potential of the buffer's input device to the buffered signal, nearly zero effective input capacitance is achieved, mitigating charge sharing of the sampled voltage. Second, trimming of $C_L$ and $C_R$ is accomplished by connecting trim capacitors to *OutBufm* instead of ground when they are not selected, boostrapping them.

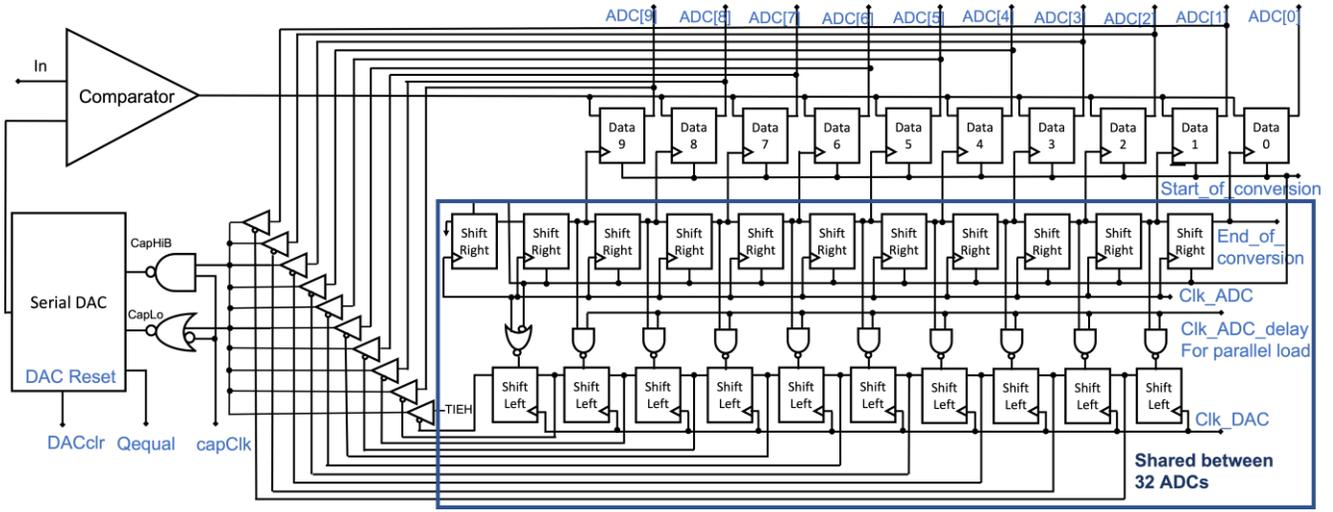

Fig. 4. Serial SAR digital control logic uses a walking one to control the common shift right, shift left and parallel load operations which is shared across 32 ADCs. Tristate buffers are used to connect the data storage register to the DAC for reusing the pixel specific ADC information.

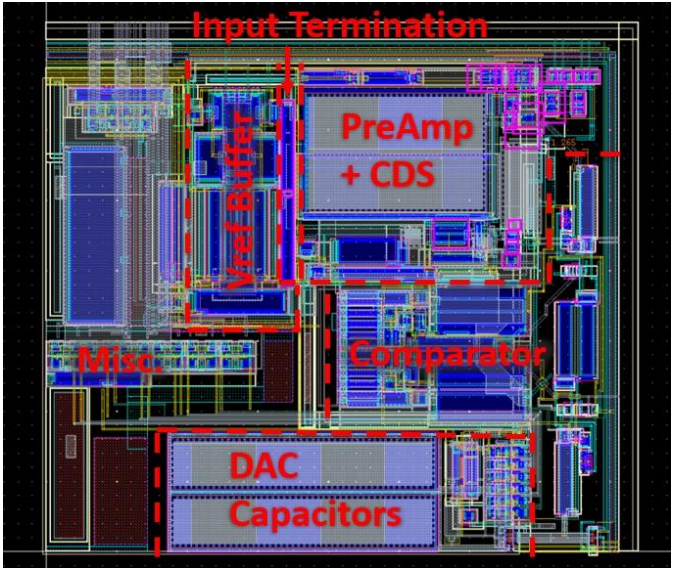

Fig. 5. Layout capture of a single SPROCKET ADC pixel

Clearly, *OutBufm* drives a much larger capacitive load than *OutBufp*. For this reason, the design of the comparator is augmented in SPROCKET by placing an additional common-drain buffer between *OutBufm* and the trim capacitors. The additional driving capability of this buffer, along with modifications to the doping of switch devices in the capacitive DAC, allow the SAR ADC to operate at 100 KSPS, approximately four times faster than the ADC in [11].

### D. Compact 100 KSPS Serial SAR ADC

The architecture of the SPROCKET ADC is a ten bit serial successive approximation register (SAR) ADC. In order to achieve ten bit resolution within the power and area constraints of an in-pixel front-end, the ADC generates the successive approximation voltages using a charge redistribution DAC based on two capacitors $C_R$ and $C_L$. Six binary weighted trim capacitors are used to tune $C_L = C_R$.

Once the capacitors are trimmed to have equal values, the ADC generates a reference voltage by briefly asserting *CapHi* or *CapLo* to charge $C_R$ either to $V_{ref}$ or to zero volts, and then asserting *Qequal* to short the positive terminals of $C_R$ and $C_L$ together [11]. The final voltage developed at the positive terminal of $C_L$ (the negative input to the comparator) after $N$ phases is given by:

$$\sum_{n=1}^{N} \frac{V_{ref} h_n}{2^{N-n+1}}$$

Where $h_n$ is 1 if CapHi is asserted on clock cycle $n$ and 0 if CapLo is asserted. As this expression suggests, the DAC requires $n$ clock cycles to compute an $n$-bit approximation.

After each approximation is computed, the ADC compares it to the sampled voltage stored across $C_{samp}$ and $C_{pre}$ to decide the next $h_n$ and thus the value of the next approximation, which will have a resolution one bit greater than the previous. DACclr is then asserted to clear the value on $C_L$. Thus one 10-bit ADC acquisition requires 55 clock cyles: $\sum_{n=1}^{10} n = 45$ cycles to compute 10 approximations of increasing precision, plus 10 DACclr cycles.

An example of CapHi, CapLo, and DACclr waveforms and the resulting voltage on $C_L$ is shown in the graph inset in Fig. 3. In this example, the sampled voltage (not shown) is near the high end of the ADC's dynamic range, and the final approximation is 10b'1111101101.

### E. Digital logic for the serial SAR ADC

A 10b serial SAR ADC typically requires a 10b data storage register, 10b DAC control register and a 10b register for the sequence and control logic [12]. This compact, low power digital logic uses only 10b register per ADC as shown in

TABLE I
SPROCKET KEY SPECIFICATIONS

| FEATURE | TARGET | RESULT | UNITS |
|---|---|---|---|
| Input referred noise | < 1 | 0.48 | $e^-_{rms}$ |
| Input dynamic rang | 9000 | 9300 | $e^-$ |
| Sample Rate | 100 | 100 | kHz |
| Temperature | 100 | 100 | K |
| Power | < 50 | 40.25 | μW |

4. The sequence and control logic which performs functions such as shift left, shift right and parallel data transfer is shared among 32 ADCs. The data storage register and the DAC control register is dependent on the amount of charge deposited per pixel and contains unique data therefore it cannot be shared across pixels. However, since the DAC control register effectively uses a subset of information in the data storage register, the circuit utilizes tristate buffers to reuse the data without duplicating the logic. Furthermore, in order to reduce the dynamic power consumption, the digital switching activity is limited to clocking only one register at a time by using a round-robin, "walking one" based architecture which sequentially selects the data to be used for the serial DAC operation as well as the register for latching the ADC output.

## III. SIMULATION RESULTS

The performance of SPROCKET has been verified using post-parasitic-extraction simulations. A Skipper CCD-in-CMOS sensor conversion gain of approximately $110 \mu V/e^-$ is based on conservative simulation results, and is used to convert voltage domain results to number of electrons. Table I summarizes the key specifications for SPROCKET and achieved results in simulation.

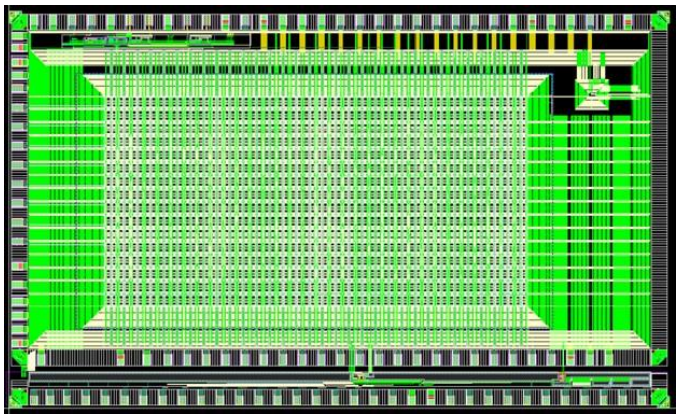

Fig. 6. Layout capture of the entire September 2022 Test Chip. The two mini-arrays occupy most of the chip area, while the test structure is at the very bottom inside a secondary pad ring.

## IV. ARRAY-LEVEL INTEGRATION OF SPROCKET

The SPROCKET front-end is monolithically integrated with a digital back-end. 2x2 clusters of SPROCKET front-ends are grouped into analog "islands" with separated deep N-wells, and the space between these N-wells is filled by synthesized digital logic. The back-end distributes digital control signals, and it contains six calibration registers per pixel for DAC trimming. It also implements the SAR operation described above and routes ADC output data to a serializer and LVDS transmitter on the peripheral for off-chip transmission. Sequential read-out of all pixels is possible, but a parallel effort is designing compression algorithms to improve read-out speed.

A test chip was submitted in September 2022 containing two 32x32 mini-arrays of SPROCKET pixels as well as a test structure for the analog front-end. A layout capture of the full test chip is shown in Fig. 6 while a single SPROCKET pixel with annotated blocks is shown in Fig. 5

In this prototype, the two compression algorithms implemented in the two mini-arrays are (1) full readout, i.e. no compression, and (2) zero-suppressed readout, in which only pixels with non-zero data are read. All control signals and reference voltages are provided off-chip, and reference currents are mirrored by an on-chip bias block to provide front-end bias voltages.

## V. TEST RESULTS

The compact serial SAR ADC design described above was prototyped in a previous design [8]. Testing demonstrated that DNL < 0.5 LSB was achievable with an increase in tuning capacitance, which was implemented in SPROCKET.

The September 2022 test chip has been received, and testing is currently underway. Although this chip will not be heterogeneously integrated to a Skipper-CCD-in-CMOS sensor, several array pixels are connected to wirebond pads, enabling single-channel testing with a separate Skipper-CCD-in-CMOS chip which has been developed by Fermilab. Test results will be presented at the conference.

## VI. CONCLUSIONS AND FUTURE WORK

The SPROCKET 65nm ASIC delivers a low-noise, low-power, area-constrained solution for the massively parallel readout of Skipper CCD-in-CMOS arrays. The design has been fabricated and is currently undergoing test.

Future SPROCKET chips that are already in design will expand the readout speed by allowing the analog pile-up of multiple samples before digitization, and by implementing more sophisticated compression algorithms. Future goals include the integration of analog references and digital control signals on-chip and expansion of the array size, with the ultimate goal of producing a full-reticle Readout IC and heterogeneously integrating it with a co-designed Skipper-CCD-in-CMOS image sensor.

REFERENCES

[1] J. Estrada, "Observatory of Skipper CCDs Unveiling Recoiling Atoms (OSCURA)," Available: https://astro.fnal.gov/science/dark-matter/oscura/, Accessed: 2021-05-01.
[2] J. Janesick, K. Klaasen, and T. Elliott, "Ccd charge collection efficiency and the photon transfer technique," in *Solid-state imaging arrays*, vol. 570. International Society for Optics and Photonics, 1985, pp. 7–19.
[3] J. Janesick, *Scientific charge-coupled devices*. SPIE press, 2001, vol. 83.

[4] G. Ferna´ndez Moroni, J. Estrada, G. Cancelo, S. E. Holland, E. E. Paolini, and H. T. Diehl, "Sub-electron readout noise in a skipper CCD fabricated on high resistivity silicon," *Experimental Astronomy*, vol. 34, no. 1, pp. 43–64, 2012. [Online]. Available: https://doi.org/10.1007/s10686-012-9298-x

[5] J. R. Janesick, T. S. Elliott, A. Dingiziam, R. A. Bredthauer, C. E. Chandler, J. A. Westphal, and J. E. Gunn, "New advancements in charge-coupled device technology: subelectron noise and 4096 x 4096 pixel CCDs," in *Charge-Coupled Devices and Solid State Optical Sensors*, vol. 1242, 1990. [Online]. Available: https://doi.org/10.1117/12.19452

[6] J. Tiffenberg et al., "Single-electron and single-photon sensitivity with a silicon skipper CCD," *Physical Review Letters*, vol. 119, no. 13, p. 131802, 2017.

[7] N. Wermes, "Pixel detectors ... where do we stand?" *Nuclear Instruments and Methods in Physics Research Section A: Accelerators, Spectrometers, Detectors and Associated Equipment*, vol. 924, pp. 44–50, apr 2019. [Online]. Available: https://doi.org/10.1016\%2Fj.nima.2018.07.003

[8] G. A. Carini *et al.*, "Hybridized maps with an in-pixel a-to-d conversion readout asic," *Nuclear Instruments and Methods in Physics Research*, vol. 935, pp. 232–238, 2019. [Online]. Available: https://www.sciencedirect.com/science/article/pii/S0168900219304334

[9] F. Fahim *et al.*, "A low-power, high-speed readout for pixel detectors based on an arbitration tree," *IEEE Trans. VLSI Syst.*, vol. 28, no. 2, pp. 576–584, 2020. [Online]. Available: https://doi.org/10.1109/TVLSI.2019.2953871

[10] H. Wey and W. Guggenbuhl, "An improved correlated double sampling circuit for low noise charge coupled devices," *IEEE Transactions on Circuits and Systems*, vol. 37, no. 12, pp. 1559–1565, 1990.

[11] T. Zimmerman, G. Deptuch, and F. Fahim, "Compact, low power, high resolution adc per pixel for large area pixel detectors." U.S. Patent 11,108,981, issued August 31, 2021.

[12] R. E. Suarez, P. R. Gray, and D. A. Hodges, "All-mos charge-redistribution analog-to-digital conversion techniques. ii," *IEEE Journal of Solid-State Circuits*, vol. 10, no. 6, pp. 379–385, 1975.